%
% Last modified: 08/11/00
%
% Page break inserted: start of section 4.
%
\def\gap {\vskip 6pt}
\def\indenton {\parindent=20pt}
\def\indentoff {\parindent=0pt}
\def\ref#1#2#3#4#5#6{[#1] #2, {\sl #3}, {\bf #4}, #5, (#6).\vskip 6pt
\par}
\magnification=\magstep1
\vsize=25 true cm
\goodbreak
\indenton
\vskip 1cm
\centerline {\bf ABSORPTION OF ENERGY AT A METALLIC SURFACE} 
\centerline {\bf DUE TO A NORMAL ELECTRIC FIELD}
\vskip 1 cm
\centerline {Michael Wilkinson}
\gap\gap
\centerline {Faculty of Mathematics and Computing,}
\centerline {Open University,}
\centerline {Walton Hall,} 
\centerline {Milton Keynes, MK7 6AA,}
\centerline {England.}
\vskip 1 cm
\centerline{\bf Abstract}
\gap
The effect of an oscillating electric field normal to a metallic
surface may be described by an effective potential. This induced
potential is calculated using semiclassical variants of the 
random phase approximation (RPA). Results are obtained for
both ballistic and diffusive electron motion, and for two 
and three dimensional systems. The potential induced within the
surface causes absorption of energy. The results are applied
to the absorption of radiation by small metal spheres and discs.
They improve upon an earlier treatment which used the Thomas-Fermi
approximation for the effective potential.
\gap\gap\gap
\vfill
\eject
%
%
%
%%%%%%%%%%%%%%%%%%%%%%%%%%%%%%%%%%%%%%%%%%%%%%%%%%%%%%%%%%%%%%%%%%%%%
%
\noindent{\bf 1. Introduction}
\gap
When an external electric field is applied to a 
metal, polarisation charges are induced on the surface.
At zero frequency, these charges create an additional
potential which exactly cancels the applied electric 
field within the interior of the metal. 
This paper discusses the form of the potential
at frequencies $\omega $ which are small compared to the 
plasma frequency of the metal, $\omega_{\rm p}$, for the
case where the external field is perpendicular
to the surface of the metal. The potential is
calculated self-consistently, using two distinct simplified 
versions of the \lq random phase approximation' (RPA) approach.
The standard random phase approximation is discussed
in [1]. The formulation of the \lq semiclassical' variants
of the RPA used in this paper was discussed in [2].

This effective potential is not directly measurable,
but it does have an influence on the electromagnetic
response of the surface, in particular on the absorption
of radiation. The results will be used
to address a long-standing problem, concerning the 
theory for absorption of radiation by small conducting
particles. 
Standard electromagnetic theory predicts that the 
absorption coefficient $\alpha(\omega)$ of a dispersion of small metal
particles is proportional to $\omega^2$, and the coefficients
are known for simple geometries such as spheres [3], or discs with
the electric field vector in the plane of the disc [4]. In
the standard treatment the conductivity of the metal
is assumed to be local, with the current density at position
${\bf r}$ proportional to the electric field at ${\bf r}$.
This assumption is valid if the motion of electrons is diffusive.
In very small particles the bulk mean free path may be larger
than the size of the particle, and in this ballistic case the conductivity 
must be treated as a non-local quantity. The frequency dependence
of the absorption coefficient has not previously received a
satisfactory treatment in the ballistic case. The final
result of this paper will be expressions for the absorption
coefficient of the form 
$$\alpha(\omega)\sim K_3\omega^2\ , \ \ \ {\rm spheres}\eqno(1.1a)$$
$$\alpha (\omega)\sim K_2\omega\ , \ \ \ {\rm discs}\ .\eqno(1.1b)$$
In the latter case, the electric field is polarised in the plane
of the disc.
 
Previous papers discussed the absorption of low frequency 
radiation in particles with ballistic electron motion using a 
Thomas-Fermi approximation: reference [5] treated spherical 
particles in three dimensions, and reference
[6] discussed discs with the electric field
polarised in the conducting plane. The surface may be 
smooth enough that the electron is reflected specularly,
in which case the motion is integrable due to conservation of 
angular momentum, or it may be a rough surface resulting
in ergodic electron motion. The earlier works [5] and [6]
consider both integrable and ergodic cases. It was shown that 
in the integrable case the absorption coefficient is the sum of 
contributions from resonant absorption by electrons
with angles of incidence $\theta $ satisfying the
condition 
$$(v_{\rm F}/a\omega)(n\pi \pm \theta_n)=\sin\theta_n
\eqno(1.2)$$
where the integer $n$ labels the resonance, $v_{\rm F}$ is the 
Fermi velocity, and $a$ is the radius. There are no
resonances below the frequency $\omega_{\rm c}=v_{\rm F}/a$,
so that there is no absorption when $\omega <\omega_{\rm c}$.
As the frequency increases, more resonances contribute,
and it was shown that when $\omega\gg \omega_{\rm c}$, the 
frequency-averaged behaviour of the sum of these 
resonances is of the form (1.1). In the case of ergodic 
electron motion there are no resonances, but (1.1) continues
to apply (although the coefficients $K_2$ and $K_3$ are 
different). The prediction that the
absorption coefficient is proportional to frequency in the
two dimensional case was a surprising result. 

In [5] and [6] it was also shown that 
when $\omega \gg \omega_{\rm c}$, the frequency-averaged 
absorption can be obtained correctly by treating the collisions 
of the electron with the surface as if they are independent 
events. The present paper adopts this simplification, which
avoids summing over contributions from a large number of 
resonances.

Reference [2] gave a comprehensive discussion of the equations
underlying the treatment of absorption of radiation, and 
concluded that the Thomas-Fermi 
approximation is not sufficient, even when the electron motion 
is ballistic. The form of the potential was shown to be very 
different from the Thomas-Fermi approximation when 
$\omega \gg \omega_{\rm c}$. The conclusions
of references [5] and [6] should therefore be re-evaluated,
particularly the prediction that $\alpha (\omega) \sim \omega$
in the two-dimensional case. This paper treats absorption by 
particles with ballistic electron motion using two distinct approximations 
to the \lq random phase approximation' (RPA), instead of the simpler 
Thomas-Fermi approximation used in [5] and [6]. The conclusions are
consistent with (1.1) (but different values for the coefficients 
$K_3$ and $K_2$ are obtained).

Two simplifications of the RPA approach are 
discussed in section 2. The first will be termed the \lq image source
approximation', and leads to a slight simplification of the RPA
equations. It is valid in the neighbourhood of a flat surface,
when the frequency is small compared to the plasma frequency. 
The second approach, previously discussed in [2], will be termed 
the \lq semiclassical' RPA method. It makes more radical assumptions, 
and leads to simpler equations.
The results of both approaches are approximate, rather than
being the leading term of an asymptotic theory. 
Section 3 obtains the self-consistent
effective potential which describes the electric field within the
surface. First this is calculated using the semiclassical RPA method
for both two and three dimensional systems, and then using the image source
approximation for the simpler case of a three dimensional system. 
Section 4 considers the transfer of energy to an electron
colliding with the surface. Again, the calculation is carried out
first for the semiclassical RPA method, and then using the image 
source approximation in the three dimensional case. 
The rate of absorption of energy is calculated in section 5, 
for both spheres and discs. Section 6 is a brief discussion
of the validity of the results.

This work uses a free electron model for the conduction electrons,
as described in [5]. The symbol $e$ will denote the magnitude of 
the electron charge, and the potential $\phi $ will denote the
potential energy of an electron (rather than the electostatic potential). 
Following common practice equations will be written as equalities, 
despite the fact that most of them are approximate relations. 
References [2] and [5] contain ample discussion of and references
to recent literature.
\gap\gap\gap
\noindent{\bf 2. Two semiclassical RPA methods}
\gap
\noindent{\sl 2.1 The RPA equations}
\gap
A polarisable medium is perturbed by applying a time-dependent
external potential, which is specified in the frequency domain 
by a function $\phi_{\rm ext}({\bf r},\omega)$. The motion
of electrons within the medium may be approximated by an 
independent particle effective Hamiltonian, which contains
an effective potential $\phi ({\bf r},\omega)$.
By analogy with Dirac notation, these potentials can 
be notated as function space vectors
$\vert \phi_{\rm ext})$ and $\vert \phi )$ respectively.
The effective potential must take account of the fact that the
medium is polarised by the externally applied field, so that
the effective potential is the sum of the external potential
and the potential $\vert \phi_{\rm pol})$ generated by the polarisation 
charge $\vert \rho )$ through the action of the Coulomb operator 
$\hat U$:
$$\phi_{\rm pol}({\bf r},\omega)\equiv ({\bf r}\vert \phi_{\rm pol})
=({\bf r}\vert \hat U \vert \rho )\equiv
{-e\over {4\pi \epsilon_0}}\int d{\bf r}'
{1\over {\vert {\bf r}-{\bf r}'\vert}}\rho ({\bf r'},\omega)
\ .\eqno(2.1)$$
The induced charge density $\vert \rho )$ is obtained from the 
effective potential $\vert \phi )$ by multiplication 
by the polarisability operator $\hat \Pi (\omega)$, i.e.
$\vert \rho)=\hat \Pi (\omega) \vert \phi )$.
Writing $\vert \phi )=\vert \phi_{\rm ext})+\vert \phi_{\rm pol})$ 
gives a single equation which should be solved for the
effective potential: the RPA equation is
$$\vert \phi )=\vert \phi_{\rm ext})+\hat U\hat \Pi (\omega) \vert \phi)
\ .\eqno(2.2)$$
A thorough discussion of the RPA method is given in [1].
\gap
\noindent{\sl 2.2 The image source approximation}
\gap
The polarisability $\hat \Pi (\omega)$ is 
related to the spatial probability propagator by
$$\hat \Pi(\omega)=e\nu \bigl[\hat I+{\rm i}\omega \hat P(\omega)\bigr]
\eqno(2.3)$$
where 
$({\bf r}\vert P(\omega) \vert {\bf r}')$ is the Fourier transform of
the probability $P({\bf r},{\bf r}',t)$
of an electron initially at ${\bf r}'$ being located at
${\bf r}$ after time $t$. A semiclassical derivation of this relation
is given in [2].

In the vicinity of a flat surface, the coordinate space representation
of the polarisability may be approximated as follows:
$$\Pi({\bf r},{\bf r}',\omega)
\equiv ({\bf r}\vert \hat \Pi(\omega)\vert {\bf r}')
=e\nu \bigl[\delta({\bf r}-{\bf r}')+{\rm i}\omega P({\bf r}-{\bf r}',\omega)
+{\rm i}\omega P({\bf r}-{\bf r}'_{\rm R},\omega)\bigr]
\ .\eqno(2.4)$$
where $P({\bf r}-{\bf r}',\omega)$ is the Fourier transform of the
free space propagator $P({\bf r}-{\bf r}',t)$, and
${\bf r}_{\rm R}$ is the reflction of the point ${\bf r}$ in
the plane of the boundary. 

Specialising to the case where the boundary is a flat plane at $z=0$,
and where the the potential $\phi (z)$ depends only on the distance 
from the boundary, the equation $\vert \rho )=\hat \Pi(\omega) \vert \phi)$ 
may be written
$$\rho(z)=e\nu \phi(z)+{\rm i}e\nu \omega \int dx'\int dy' \int dz'\,
\bigl[P((x',y',z-z'),\omega)+P((x',y',z+z'),\omega)\bigr]\phi(z')
\ .\eqno(2.5)$$
The integrals will be taken to be over all space, with $\phi(z)=0$
for $z<0$. Equation (2.5) will be termed the \lq image source
approximation'.

In view of the isotropy of the free space propagator, its Fourier
transform $\tilde P({\bf k},\omega)$ is a function of 
the magnitude $k=\vert {\bf k}\vert$ of the wavevector:
$$p(k,\omega)= \tilde P({\bf k},\omega) \equiv
\int d{\bf r}\ \exp[{\rm i}{\bf k}.{\bf r}]\,P({\bf r},\omega)
\ ,\ \ \ k\equiv \vert {\bf k}\vert
\ .\eqno(2.6)$$
\par
The function $p(k,\omega)$
can be determined by Fourier transformation of a semiclassical 
approximation of the position representation of the free space
propagator, valid when $k\ll k_{\rm F}$. In [2] it was shown
that the resulting Fourier representation of the free space 
polarisability may be expressed in terms of a single scaling 
variable $\lambda$:
$$\Pi(k,\omega)=e\nu \bigl[1+{\rm i}\omega p(k,\omega)\bigr]
=e\nu \bigl[1+g(\lambda)\bigr]
\ .\eqno(2.7)$$
The dimensionless variable $\lambda$ takes different 
forms for ballistic or diffusive electron dynamics
$$\lambda=\biggl\{
{kv_{\rm F}/\omega\ , \ \ \ {\rm ballistic}
\atop{
k\sqrt{D/\omega}\ ,\ \ \ {\rm diffusive}
}}
\eqno(2.8)$$
where $D$ is the diffusion constant.
In the ballistic case the form of the function $g(k)$ depends
upon the dimensionality of space. In two dimensions:
$$g_2(\lambda)=\biggl\{
{-(1-\lambda^2)^{-1/2}\ ,\ \ \lambda<1
\atop{
{\rm i}(\lambda^2-1)^{-1/2}\ ,\ \ \lambda>1
}}
\eqno(2.9)$$
and in three dimensions
$$g_3(\lambda)=-{1\over{2\lambda}}
\log_{\rm e} \bigg\vert {\lambda+1\over {\lambda-1}}\bigg\vert
+{\pi{\rm i}\over{2\lambda}}\Theta (\lambda-1)
\eqno(2.10)$$
(where $\Theta(x)$ is the Heaviside function, with increasing
unit step at $x=0$).
The limiting forms of $g(\lambda)$ for small and large argument
are informative:
$$g_d(\lambda)=-\bigl(1+{1\over d}\lambda^2\bigr)+O(\lambda^3)
\eqno(2.11)$$
$$\lim_{\lambda \to \infty}\bigl[\lambda g_d(\lambda)\bigr]
=\biggl\{
{{\rm i}\ ,\ \ \ (d=2)
\atop{
{\pi\over 2}{\rm i}\ ,\ \ \ (d=3)
}}
\ .\eqno(2.12)$$
In both two and three dimensional ballistic motion, $g_d(\lambda)$ has an
integrable divergence at $\lambda =1$.
In the case of diffusive electron motion, the form of the function 
$g(\lambda)$ is the same in two and three dimensions:
$$g_{\rm D}(\lambda)={-1\over{1+{\rm i}\lambda^2}}
\ .\eqno(2.13)$$
\gap
\noindent{\sl 2.3 The semiclassical RPA approximation}
\gap
Formally, the effective potential may be determined simply
by inverting (2.2): 
$\vert \phi )=[\hat I-\hat U \hat \Pi(\omega)]^{-1}
\vert \phi_{\rm ext})$. This approach can be implemented
numerically by expanding functions in a suitable basis set and
inverting matrices numerically. It is desirable to have  
a simpler approach which allows further analytical progress.
In [2], a \lq semiclassical' variant of the RPA approach
was introduced. It is assumed that 
(provided $\omega\ll \omega_{\rm p}$) the polarisation
charge $\vert \rho)$ is that which would be predicted by 
classical electrostatic theory (the justification for this 
is discussed in [2]). Under this assumption, the 
effective potential satisfies a much simpler equation
$$\vert \rho_{\rm cl})=\hat \Pi (\omega) \vert \phi )
\eqno(2.14)$$
where $\vert \rho_{\rm cl})$ is the classical charge distribution
function resulting from a static external field, which is assumed 
to be known. The task of determining the effective potential
is then reduced to the simpler task of inverting $\hat \Pi (\omega)$.
Calculation of $\vert \rho_{\rm cl})$ is still a difficult problem, but
solutions are obtained in various geometries in textbooks such as [3].

The form of the surface charge density is different in two and three
dimensions. In three dimensions the charge density is confined to
a narrow layer at the surface of the conductor, with thickness equal 
to the Thomas-Fermi screening length. This can be approximated by a 
delta-function distribution, a tiny distance $\varepsilon$ inside the 
surface: 
$$\rho (z)=q\,\delta (z-\varepsilon)\eqno(2.15)$$
where $q$ is the surface charge density induced by the 
externally applied fields. In (2.15) the coefficient
$q$ depends upon the position ${\bf s}$ on the surface 
of the conducting particle. 
In the case of a two-dimensional conductor in three dimesional
space, there is an inverse square-root divergence of the charge 
density at the surface of the particle:
$$\rho (z)={C\over {\sqrt{z}}}
\ .\eqno(2.16)$$
The reasons for this behaviour are discussed in [6].
\gap\gap\gap
\noindent {\bf 3. Calculation of the surface potential}
\gap
\noindent{\sl 3.1 Method for solving the semiclassical RPA equation}
\gap
Equation (2.5) will now be expressed in a purely one-dimensional form.
To this end, define $F(z,\omega)$ as the inverse Fourier transform of
$p(k,\omega)$
$$p(k,\omega)\equiv \int_{-\infty}^\infty dz\ \exp[{\rm i}kz]\, 
F(z,\omega)
\ .\eqno(3.1)$$
and note that
$$\int_{-\infty}^\infty dx' \int_{-\infty}^\infty dy' 
\int_{-\infty}^\infty dz'\, P((x',y',z-z');\omega)\,\phi(z')$$
$$=-{1\over{(2\pi )^3}}\int d{\bf r}''\, \phi (z-z'')
\int d{\bf K}\ \exp[-{\rm i}{\bf K}.{\bf r}'']\, 
\tilde P({\bf K};\omega)$$
$$=-{1\over{2\pi}}\int_{-\infty}^\infty dz''\, \phi(z-z'')\int d{\bf K} 
\ \tilde P({\bf K},\omega)\, \delta (K_x)\, \delta (K_y) 
\, \exp [-{\rm i}{\bf K}.{\bf r}'']$$
$$=-{1\over{2\pi}}\int_{-\infty}^\infty dz''\, \phi(z-z'')
\int_{-\infty}^\infty dk\ \exp[-{\rm i}kz''] 
\, p(k,\omega)$$
$$=\int_{-\infty}^\infty dz\ F(z-z',\omega)\, \phi(z')
\ .\eqno(3.2)$$
Using this result, the integrals over $x'$ and $y'$ in equation (2.5) 
may be eliminated, giving
$$\rho(z)=e\nu \biggl[\phi(z)+{\rm i}\omega 
\int_{-\infty}^\infty dz'\ 
\bigl[F(z-z',\omega)+F(z+z',\omega)\bigr]\phi(z')\biggr]
\eqno(3.3)$$
Comparison of (3.3) with (3.1) shows that $F(z,\omega)$ is related
to the inverse Fourier transform of the function $g(\lambda)$, introduced
in (2.7). In the ballistic case:
$$F(z,\omega)={1\over{2\pi {\rm i}\omega}}\int_{-\infty}^\infty
dk\ \exp[-{\rm i}kz]\,g(kv_{\rm F}/\omega)
\equiv {1\over{{\rm i}v_{\rm F}}}G(z\omega/v_{\rm F})
\eqno(3.4)$$
where $G$ is the inverse Fourier transform of $g$.
Equation (3.3) can now be written in the scaled form
$$\rho(z)=e\nu \biggl[\phi(z)+{1\over \Lambda}\int_0^\infty dz'\ 
\bigl[G\bigl({z-z'\over{\Lambda}}\bigr)
+G\bigl({z+z'\over{\Lambda}}\bigr)\bigr]\phi(z')\biggr]
\eqno(3.5)$$
where the scale length $\Lambda$ is
$$\Lambda=\biggl\{
{v_{\rm F}/\omega\ , \ \ \ {\rm ballistic}
\atop{
\sqrt{D/\omega}\ ,\ \ \ {\rm diffusive}
}}
\ .\eqno(3.6)$$
Defining a scaled distance by $x=z/\Lambda$, and a scaled charge 
density $f(x)$ and potential $\psi (x)$ by 
$$\rho(z/\Lambda)=e\nu f(\Lambda x)\ ,\ \ \ \phi(z)=\psi(\Lambda x)
\eqno(3.7)$$
equation (3.5) may be expressed in the dimensionless form
$$f(x)=\psi(x)+\int_0^\infty dx'\ \bigl[G(x-x')+G(x+x')\bigr]\psi(x')
\ .\eqno(3.8)$$
\par
Now consider how to solve equation (3.8) for the scaled potential
$\psi(x)$, given the scaled charge density $f(x)$. The function $f(x)$
is defined only for $x>0$. Also, the behaviour of $\psi(x)$
for $x<0$ is irrelevant to the form of $f(x)$ in the region
where this is defined. We may define a symmetric extension of the
function $f(x)$:
$$f_{\rm s}(x)=f(\vert x\vert)
\ .\eqno(3.9)$$
Consider a function $\psi_{\rm s}(x)$ which satisfies
$$f_{\rm s}(x)=\psi_{\rm s}(x)
+\int_{-\infty}^\infty dx'\, G(x-x')\,\psi_{\rm s}(x')
\ .\eqno(3.10)$$
This function must be symmetric: $\psi_{\rm s}(x)=\psi_{\rm s}(-x)$.
This symmetric solution satisfies the same equation as $\psi(x)$ 
(equation (3.8)) for $x<0$. We therefore solve the simpler equation
(3.10), and drop the subscript ${\rm s}$ labelling the solution.
The Fourier transform of the solution is
$$\tilde \psi(k)={\tilde f_{\rm s}(k)\over {1+g(k)}}
\ .\eqno(3.11)$$
The solution may also be expressed as a convolution
$$\psi(x)=\int_{-\infty}^\infty dx'\, K(x-x')\,f_{\rm s}(x')
\eqno(3.12)$$
where $K(x)$ is the inverse Fourier transform of $(1+g(k))^{-1}$.

These expressions require the form of the function $f_{\rm s}(x)$.
Referring to (2.15) (and taking $\varepsilon \to 0$), in three 
dimensions the Fourier tansform of the 
scaled and symmetrised charge density is, in the ballistic case 
$$\tilde f_{\rm s}(k)={2q\omega\over{e\nu v_{\rm F}}}
\ .\eqno(3.13)$$
The factor of $2$ appears in (3.13) because both the delta
function and its symmetric image contribute. From (2.16),
in two dimensions the analogous quantity is
$$\tilde f_{\rm s}(k)={2\sqrt{2\pi}C\over{e\nu}}
\sqrt{\omega\over{v_{\rm F}\vert k\vert}}
\ .\eqno(3.14)$$
\gap
\noindent{\sl 3.2 Solution of the image source approximation}
\gap
It is instructive to compare the solution of the \lq semiclassical'
RPA equation, (2.14), with that of the image source approximation, (2.5).
This is difficult in the two-dimensional case, but quite 
straightforward in the case of a flat metallic surface in 
three dimensions. The charge density is still assumed to be
a function of $z$ alone, and is related to the effective
potential $\phi(z)$ by Poisson's equation:
$${1\over{e}}{d^2\phi \over{dz^2}}={\rho(z)\over{\epsilon_0}}
\eqno(3.15)$$
so that the charge density in (3.5) may be replaced by a term
proportional to the second derivative of $\phi(z)$. 
(This simplification is not possible in two dimensions, because
it that case $\rho$ also depends upon the second derivative with respect to
the coordinate perpendicular to the conducting plane.) The resulting 
equation can be expressed in a scaled form analogous to (3.8),
and is then transformed into an equation for
the symmetrised potential, corresponding to (3.10). Here it is necessary
to note that the symmetrised potential can have a discontinuity
at $z=0$, without the charge density having a singularity there.
The electric field approaches a constant value as $z\to \infty$,
and provided $\omega\ll \omega_{\rm p}$ this field is much smaller
than the externally applied field. If the internal field is
neglected, the potential $\phi (z)$ may be assumed to approach
zero as $z\to \infty$, and as $z\to 0_+$ the slope $d\phi/dz$ 
approaches $-eq/\epsilon_0$, where $q$ is the 
integral of $\rho(z)$, i.e. the total charge per unit 
area bound to the surface. The symmetrised potential therefore
has a discontinuity of slope equal to $-2qe/\epsilon_0$ at 
$z=0$. The scaled potential $\psi_{\rm s}(x)$ satisfies
$${\epsilon_0\over{e^2\nu \Lambda^2}}{d^2\psi_{\rm s}\over{dx^2}}
=\psi_{\rm s}(x)+\int_{-\infty}^\infty dx'\ G(x-x')\,\psi_{\rm s}(x')
-{2q\over{e\nu \Lambda}}\delta (x)
\ .\eqno(3.16)$$
This version of the RPA equation can also be solved directly by a Fourier 
transform approach. Noting that for ballistic electron motion 
in three dimensions 
$\epsilon_0/e^2\nu \Lambda^2=\omega^2/3\omega^2_{\rm p}$, 
the Fourier transform of the image source approximation solution is
$$\tilde \psi_{\rm s}(k)={2q\omega\over{e\nu v_{\rm F}}}
{1\over{1+g_3(k)+{\omega^2\over {3\omega^2_{\rm p}}}k^2}}
\ .\eqno(3.17)$$
This solution is indeed very close to the semiclassical
RPA solution when $\omega \ll \omega_{\rm p}$.

It should be emphasised that these forms for the effective
potential are applicable only when the scale length $\Lambda$
is small compared to the characteristic dimension $a$ of the system.
In the low frequency limit, $\omega \ll \omega_{\rm c}$
the scale size $\Lambda$ exceeds the size of the particle. 
In this limit a Thomas-Fermi approximation should be used 
for the effective potential. In this case, the form for the 
Fourier transform of the effective potential is obtained from (3.11)
or (3.17) by replacing the function $g(k)$ by zero.
\gap
\noindent{\sl 3.3 Components of the effective surface potential}
\gap
Reference [4], which considered the case where electron motion
is diffusive, suggested writing the effective potential as the sum
of two terms, namely a \lq static' potential, which is given by the 
Thomas-Fermi approximation, and a \lq dynamic' potential, which is
proportional to frequency, and which is required to move the
static polarisation into place when the external electric field
changes. Reference [2] showed that an additional component
must be present in the ballistic case. It is interesting
to see how these components are represented in equations
(3.11), (3.13) and (3.17). The discussion will be restricted to
the three dimensional case.

Thomas-Fermi theory predicts that the potential is proportional
to the charge density: $\phi(z)=\rho(z)/e\nu$. This is equivalent
to setting $g(k)=0$ in (3.11) or (3.17). The resulting potential
will be termed the \lq static' potential, $\phi_{\rm stat}(z)$. 
In the case of the semiclassical RPA method, it is simply a 
delta function localised at the surface, and in the case
of the image source approximation it decays rapidly as a function
of distance from the boundary, with decay length 
$\lambda_{\rm s}=\sqrt{\epsilon_0/e^2\nu}$.

In the vicinity of a surface which accumulates a 
polarisation charge $q$, the internal electric field has magnitude
${\cal E}_{\rm int}={\rm i}\omega q/\sigma(\omega)$ where 
$\sigma(\omega)$ is the bulk conductivity. Reference [2] showed
that this expression for the internal field is also applicable
in the ballistic case, when the distance from the boundary
is greater than $\Lambda =v_{\rm F}/\omega$. In the ballistic case,
the bulk conductivity is $\sigma(\omega)=Ne^2/{\rm i}m\omega$, 
where $N$ is the density of conduction electrons, and in the 
diffusive case $\sigma=e^2\nu D$ for frequencies small compared to 
the collision rate. There is therefore a component of the 
symmetrised potential which is proportional to $\vert x\vert$.
In the ballistic case, this component of the scaled potential
is
$$\psi_{\rm dyn}(x)={{\rm i}e\omega q \Lambda \over {\sigma(\omega)}}
\vert x \vert
=-{3q\omega \over{2e\nu v_{\rm F}}}\vert x\vert 
\ .\eqno(3.18)$$
The generalised Fourier transform of $\vert x\vert$ is $-2/k^2$.
Consistency with (3.18) therefore requires that the Fourier
transform of $\psi_{\rm s}(x)$ approaches 
$3q\omega/e\nu v_{\rm F}k^2$ as $k\to 0$. Using equation (2.11)
to evaluate the limit of (3.11) or (3.17) as $k\to 0$ verifies
this relation.

The full effective potential may be written as the sum of 
$\phi_{\rm stat}(z)$, $\phi_{\rm dyn}(z)$, and an additional
term \lq surface' term $\phi_{\rm surf}(z)$, which decays with
a characteristic length scale $\Lambda=v_{\rm F}/\omega$.
The three components of the effective potential are illustrated
schematically in figure 1. The surface 
potential for the three dimensional ballistic case is 
$$\psi_{\rm surf}(x)={2q\omega\over {e\nu v_{\rm F}}}
\int_0^\infty dk\ \exp ({\rm i}kx)\biggl[ 
{1\over
{1-{1\over {2k}}\log_{\rm e}\vert {k+1\over{k-1}}\vert+{{\rm i}\pi\over {2k}}
\Theta(k-1)+{\textstyle{1\over 3}}({\omega\over {\omega_{\rm p}}})^2}}
-1+{3\over {k^2}}\biggr]
\eqno(3.19)$$
This expression diverges logarithmically as $x\to 0$ when 
${\omega\over{\omega_{\rm p}}}=0$.

In the case of diffusive electron motion it was shown in [7]
that the effective potential is the sum of the static and dynamic
components only. It is instructive to see how this conclusion is 
confirmed using the results of the present paper. In the diffusive 
case, it follows from (2.15) that the Fourier transform 
of the charge density is $\tilde \rho_{\rm s}(k)={2q/{e\nu}}$. 
Using (2.13) and (3.11), the Fourier transform of the effective 
potential is therefore
$$\tilde \phi_{\rm s} (k)={2q\over {e\nu}}
{1\over{1+g_{\rm D}(\sqrt{D/\omega}k)}}
={2q\over{e\nu}}\biggl[ 1-{{\rm i}\omega\over{Dk^2}}\biggr]
\ .\eqno(3.20)$$
The first term in the final bracket Fourier transforms into a delta
function, and therefore represents the static potential. The term
proportional to $\omega/ k^2$ Fourier transforms to a term
proportional to $\omega \vert z\vert $, and represents 
the dynamic potential. The additional surface potential
component is therefore absent in the diffusive case.
\gap\gap\gap
\vfill
\eject
\noindent{\bf 4. Energy transferred on collision with surface}
\gap
The objective is to calculate the energy transferred to an
electron from the externally applied electromagnetic field as it 
collides with the surface. From this point 
onwards, the discussion is specific to ballistic electron
dynamics. The problem will first be treated classically, then
this will be compared with the results of a quantum mechanical
calculation.
\gap
\noindent{\sl 4.1 Classical treatment}
\gap
Classically, the energy transfer is determined by
separate contributions from collisions with the surface
of the particle, provided the surface potential is localised
at the surface of the particle. If the electron collides with 
the surface at time $t_0$, and is in the vicinity of the surface 
for a time $\Delta t$, the energy transferred at this collision is 
$$\delta E=\int_{t_0-\Delta t}^{t_0+\Delta t} dt\  
{\partial \phi \over {\partial t}}\bigl({\bf r}(t),t\bigr)
\eqno(4.1)$$
where ${\bf r}(t)$ is the trajectory of the electron, and where
the potential $\phi ({\bf r},t)$ is
$$\phi (z,t)={\rm Re} \biggl[\exp({\rm i}\omega t)
\,\psi\bigl(z(t)/\Lambda\bigr)\biggr]
\ .\eqno(4.2)$$
For an electron incident at an angle $\theta $ from the normal
to the surface, the distance from the surface at time $t$ is
$$z(t)=v_{\rm F}\cos \theta \vert t-t_0\vert
\ .\eqno(4.3)$$
Using the fact that $\psi(x)=\phi(x\Lambda)$ is symmetric about $x=0$, the 
energy tranferred is therefore given by
$$\delta E=-\omega \ 
{\rm Im}\biggl[ \int_{-\infty}^\infty dt\ 
\exp({\rm i}\omega t)\,
\psi\bigl(v_{\rm F}\cos \theta (t-t_0)/\Lambda\bigr)\biggr]
\ .\eqno(4.4)$$
This may be expressed in terms of the Fourier transform
of $\psi$:
$$\delta E={-\Lambda\over{v_{\rm F} \cos \theta}}{\rm Im}\ 
\biggl[\exp({\rm i}\omega t_0) 
\int_{-\infty}^\infty dx\ \exp({\rm i}x\omega \Lambda/v_{\rm F}\cos \theta)
\,\psi(x)\biggr]$$
$$={-1\over{\cos \theta}}{\rm Im}\biggl[\exp({\rm i}\omega t_0)
\tilde \psi (1/\cos \theta)
\biggr]
\ .\eqno(4.5)$$
Using (3.11), this result may be expressed in terms of the 
Fourier transform of $\psi (z)$ as follows:
$$\delta E={-1\over {\cos \theta}}{\rm Im}\biggl[\exp({\rm i}\omega t_0)
{\tilde f_{\rm s}(1/\cos \theta)\over {1+g_2(1/\cos \theta)}}\biggr]
\ .\eqno(4.6)$$
\par
Now consider the form of $\delta E$ for collision of a 
ballistic electron with the surface. In the two dimensional
case, combining (4.6), (2.9) and (3.14) gives
$$\delta E={-C\over{e\nu}}\sqrt{2\pi \omega\over{v_{\rm F}}}
{\rm Im}\biggl[{\exp({\rm i}\omega t_0)
\over{\sqrt{\cos \theta}(1+{\rm i}\cot \theta)}}\biggr]
\eqno(4.7)$$
which may be written
$$\delta E={C\over{e\nu}}\sqrt{2\pi \omega\over{v_{\rm F}}}
{\sin \theta\over{\sqrt{\cos \theta}}} {\rm Re}
\biggl[\exp\bigl[{\rm i}\bigl(\omega t_0+\chi_2(\theta)\bigr)\bigr]\biggr]
\eqno(4.8)$$
where $\chi_2(\theta)$ is a real-valued phase.
In the three dimensional case the analogous expression, obtained
using (2.10) and (3.12), is
$$\delta E={2q\omega\over{e\nu v_{\rm F}}}{\rm Im}
\biggl[{\exp({\rm i}\omega t_0)
\over{\cos \theta \bigl(1+g_3(1/\cos \theta)\bigr)}}\biggr]
={2q\omega\over{e\nu v_{\rm F}}}S(\theta) {\rm Re}
\biggl[\exp \bigl[{\rm i}
\bigl(\omega t_0+\chi_3(\theta)\bigr)\bigr]\biggr]
\eqno(4.9)$$
where 
$$S(\theta)={1\over{\cos \theta}}\bigg\vert
{1\over{1-{\textstyle{1\over 2}}\cos \theta 
\bigl[\log_{\rm e}(1+\cos \theta)-\log_{\rm e}(1-\cos \theta)\bigr]
+{\pi \over 2}{\rm i}\cos \theta}}\bigg\vert
\eqno(4.10)$$
and $\chi_3(\theta)$ is a phase.
Both the two-dimensional and the three-dimensional expressions 
diverge at grazing incidence, where $\theta \to \pi/2$. Physically,
this can be interpreted in terms of the wavepacket spending a long 
time in contact with the barrier for a reflection at grazing 
incidence. A quantum mechanical treatment will take account of 
the fact that the wavefunction approaches zero at the surface.
This removes the divergence.
\gap
\noindent{\sl 4.2 Quantum treatment}
\gap
Now consider how the expression (4.6) must be modified to
take account of quantum mechanics. The approach will be to
consider the collision of a wavepacket with the surface. The 
expectation value of the energy transferred will be calculated:
this is 
$$\langle \delta E \rangle=\int_{-\infty}^\infty dt
\int_0^\infty dz\ P(z,t)\, {\partial \phi\over{\partial t}}(z,t)
\eqno(4.11)$$
where $P(z,t)=\vert \Psi(z,t)\vert^2$ is the probability density
for the electron to be at a distance $z$ from the surface. If
the wavepacket can be chosen to be sufficiently well localised,
the energy transfer can be assumed to be equal to this expectation
value. On the other hand, if the length scale $L$ over 
which the wavepacket is 
localised is too large, then the expectation value will represent
an average over the temporal variation of the potential, and
$\langle \delta E\rangle$ will under-estimate the magnitude of the 
energy transferred. The criterion is $L\ll v_z/\omega$, where
$v_z=v_{\rm F}\cos \theta$ is the speed at which a Fermi-surface 
electron with angle of incidence $\theta$ approaches the surface.
If the energy of the electron is to be well defined (and close
to the Fermi level), then $k_{\rm F}L\ll 1$, where $k_{\rm F}$
is the Fermi wavevector. These two inequalities for $L$ are
compatible provided $\hbar \omega \ll E_{\rm F}$, which is assumed
throughout.

The wavefunction of the wavepacket which collides with the surface
at time $t_0=0$ may be written in the approximate form
$$\Psi(z,t)=\exp(-{\rm i}E_{\rm F}t/\hbar)
\bigl[\exp({\rm i}p_zz/\hbar)f(z-v_zt)
-\exp(-{\rm i}p_zz/\hbar)f(z+v_zt)\bigr]\eqno(4.12)$$
where $f(x)$ is a symmetric function, which decays rapidly when
$\vert x\vert \gg L$. Dispersion of the wavepacket is unimportant,
and is ignored in writing (4.12). The function $f$ is normalised, and its 
autoconvolution $F$ is required:
$$F(x)=\int_{-\infty}^\infty dx'\ f(x-x')f(x')\ ,\ \ \ F(0)=1
\ .\eqno(4.13)$$
Assuming that $\delta E=\langle \delta E\rangle$, and substituting
(4.12) into (4.11) gives
$$\delta E={\textstyle{1\over 2}}\int_{-\infty}^\infty dz
\int_{-\infty}^\infty dt\ 
{\partial \phi \over {\partial t}}(z,t)
\bigl[f^2(z-v_zt)+f^2(z+v_zt)-2\cos(2p_zz/\hbar)f(z-v_zt)f(z+v_zt)\bigr]
\ .\eqno(4.14)$$
The integral over time will be performed first. It has been assumed that
the envelope function varies much more rapidly than the variation in 
time of the potential $\phi (z,t)$, although the potential may vary
rapidly as a function of $z$. Using (4.13), the energy transfer may 
therefore be approximated as follows:
$$\delta E={1\over {2 v_z}}\int_{-\infty}^\infty dz\ 
{\partial \phi \over {\partial t}}(z,z/v_z)+
{\partial \phi \over {\partial t}}(z,-z/v_z)
-2\cos(2p_zz/\hbar)F(2z)
{\partial \phi \over {\partial t}}(z,0)
\ .\eqno(4.15)$$
If the potential varies sinusoidally in time, such
that $\phi(z,t)={\rm Re}[\exp({\rm i}\omega t)\phi(z)]$,
then (4.15) becomes
$$\delta E={-\omega\over {v_{\rm F}\cos \theta}} {\rm Im}\biggl[
\int_{-\infty}^\infty \!\! dz\ \exp({\rm i}\omega z/v_{\rm F}\cos \theta)
\phi (z)-F(2z)\phi(z)\cos(2mv_{\rm F}z\cos \theta/\hbar)\biggr]
\ .\eqno(4.16)$$
The first term of (4.16) is a Fourier transform of $\phi(z)$.
Comparison with (3.17) shows that this takes the form
$$\tilde \phi (k)={2q\over{e\nu}}\biggl[
{1\over{1+g_3(k v_{\rm F}/\omega)+{\epsilon_0\over{e^2\nu}}k^2}}
\biggr]
\ .\eqno(4.17)$$
The second term of (4.16) is the Fourier transform of the
product $F(2z)\phi(z)$ evaluated at $k=2mv_{\rm F}\cos \theta/\hbar$.
This may be obtained by convolution of (4.17) with the Fourier
transform of $F(2z)$. The function (4.17) has support 
$\sqrt{e^2\nu /\epsilon_0}=1/\lambda_{\rm s}$, where
$\lambda_{\rm s}$ is the Thomas-Fermi screening length, and
has a structure close to $k=0$ associated with the function $g$, 
with a narrower support, $\omega/v_{\rm F}=1/\Lambda$. 
The assumptions concerning the support $L$ of $F(z)$ imply 
that $\Lambda \gg L \gg \lambda_{\rm s}$. When calculating
the Fourier transform of $\phi(z)F(2z)$ using the convolution
theorem, the structure in $\tilde \phi (k)$ associated with
the function $g$ is suppresed, because the support of $\tilde F(k)$
is broader. The function $\tilde \phi(k)$ remains unchanged
in other respects, because the support of $\tilde F(k)$ is narrower
than the overall support of $\tilde \phi (k)$. 
The quantum mechanical expression for the energy transferred,
calculated using the image source approximation for the effective
potential, can therefore be approximated as follows
$$\delta E={-2q\omega\over{e\nu v_{\rm F}\cos \theta}}{\rm Im}\biggl[
{1\over {1+g_3(kv_{\rm F}/\omega)}}-
{1\over{1+{4\epsilon_0 m^2 v_{\rm F}^2\over{\nu e^2\hbar^2}}\cos^2\theta}}
\biggr]
\ .\eqno(4.18)$$
This simplifies to
$$\delta E={2q\omega\over{e\nu v_{\rm F}}}{\rm Im}
\biggl[{\cos \theta -\Gamma g_3(1/\cos \theta)/\cos\theta
\over{\bigl(\Gamma +\cos^2 \theta\bigr)
\bigl(1+g_3(1/\cos \theta)\bigr)}}\exp({\rm i}\omega t_0)\biggr]
\eqno(4.19)$$
where the time $t_0$ of collision with the surface has been inserted,
and where
$$\Gamma={e^2\nu\hbar^2\over{4\epsilon_0m^2v_{\rm F}^2}}
\ .\eqno(4.20)$$
The constant $\Gamma$ was introduced in [5]. It may be expressed
in the alternative forms
$$\Gamma={2\lambda_{\rm F}\over{\pi a_0}}
={2^{5/3}\over{3^{2/3}\pi^{4/3}}}{r_{\rm s}\over {a_0}}
\eqno(4.21)$$
where $a_0$ is the effective Bohr radius, and $r_{\rm s}$
is the radius of a sphere containing a single electron. 
Equation (4.19) will also be written in the form
$$\delta E= 
{2q\omega\over{e\nu v_{\rm F}}}S(\theta)\cos[\omega t_0+\chi_3(\theta)]
\eqno(4.22)$$
where
$$S(\theta)=\bigg\vert {\cos \theta-\Gamma g_3(1/\cos \theta)/\cos \theta
\over{(\Gamma+\cos^2 \theta)\bigl(1+g_3(1/\cos \theta)\bigr)}}\bigg\vert
\eqno(4.23)$$
The formulae
above should reduce to the Thomas-Fermi theory when the function $g$ is
replaced by zero, and (4.20) does indeed reduce to equation (3.25) of 
reference [5] upon setting $g=0$. 

The divergence of (4.10) at glancing incidence
is absent in this more sophisticated quantum mechanical treatment, 
because the wavefunction of the electron
vanishes at the surface, and the electron is therefore unable to be
influenced by the potential there. Equation (4.19) approaches
the prediction from (4.9) and (4.10) in the limit $\Gamma \to 0$ for
all values of $\theta $ except $\pi/2$, because as $\Gamma \to 0$
the Fermi wavelength becomes small compared to the Thomas-Fermi
screening length, and the electron is able to penetrate closer
to the surface.

In figure 2 the energy transferred to a reflected electron is plotted 
as a function of the angle of incidence $\theta $, for two different
values of $\Gamma $, and for the semiclassical RPA, which corresponds 
to the limit $\Gamma \to 0$.
\gap\gap\gap
\noindent {\bf 5. Rate of absorption of energy}
\gap
\noindent{\sl 5.1 General considerations}
\gap
The absorption coefficient of a suspension of small 
particles is determined by the rate at which an
individual particle absorbs energy. A semiclassical approach
developed in earlier papers [5,6,2] shows how the rate of absorption
of energy by the electron gas may be expressed in terms of the 
variance of the change of energy of a single electron.
For non-interacting fermions, the rate of change of the total
energy $E_{\rm T}$ of the electron gas may be written
$${dE_{\rm T}\over {dt}}=2V\nu D_E\eqno(5.1)$$
where $V$ is the volume of the particle, and $D_E$ is a diffusion
constant for single electron energies. The factor of $2$ in (5.1) 
accounts for spin degeneracy, and $\nu$ is interpreted as the
density of states per unit volume per spin. The diffusion constant 
is defined by writing
$$\langle \Delta E^2(t)\rangle=2D_Et\eqno(5.2)$$
where $\Delta E(t)$ is the energy transferred to an
electron after time $t$.

In the context of this paper $\Delta E(t)$ is the sum of the 
energy $\delta E_j$ transferred on collisions of the electron with the 
surface:
$$\Delta E(t)=\sum_j \delta E_j\eqno(5.3)$$
where the sum runs over all collisions between times $0$ and $t$.
The variance in (5.2) is defined in terms of a phase space
average for electrons at the Fermi energy:
$$\langle \Delta E^2(t)\rangle ={
\int d\alpha\ \Delta E^2(\alpha) \, \delta (H(\alpha)-E_{\rm F})
\over{
\int d\alpha\ \delta (H(\alpha)-E_{\rm F})
}}\eqno(5.4)$$
where $\alpha=({\bf q},{\bf p})$ are phase space coordinates
of an electron, and $\Delta E(\alpha)$ is the energy transferred
to an electron which is initially at $\alpha$.

In the ergodic case, 
$$\langle \Delta E^2(t)\rangle = R\, \langle \delta E^2 \rangle \, t
\eqno(5.5)$$
where $R$ is the rate of collision of particles with the boundary.
The total distance travelled by the electron in time $t$ is $v_{\rm F}t=
N\langle L\rangle$, where $\langle L \rangle$ is the mean distance
travelled between each of the $N$ collisions. The rate of collisions 
is therefore
$$R={v_{\rm F}\over{\langle L\rangle}}
\ .\eqno(5.6)$$
\gap
\noindent{\sl 5.2 Absorption by conducting discs}
\gap
The simpler case in which to evaluate the phase space average
(5.4) is for two-dimensional discs, and this will be discussed
in some detail to illustrate the approach. The calculation must be
confined to the simpler semiclassical RPA approximation, 
because the more precise image source approximation has
not been calculated in the two-dimensional case.
The integrable case will be considered first, followed by 
the ergodic case (which applies when the surface of the disc is rough).

For the case of integrable electron motion in a disc, the angular
momentum $J$ is a conserved quantity, and the coordinates
$(E,J,t_0,\varphi_0)$ are a canonical set, where $E$ is the 
energy, $t_0$ the time since the most recent collision,
and $\varphi_0$ is the polar angle of the most recent collision.
The angular momentum and the period $\tau $ between collisions
are both related to the angle of incidence, $\theta $:
$$J=mav_{\rm F}\sin \theta\ , \ \ \ \tau={2a\over{v_{\rm F}}}\cos \theta
\eqno(5.7)$$
where $a$ is the radius of the disc.
Also, the charge density induced on a disc by an electric 
field ${\cal E}$ in the plane of the disc is
$$\rho(r,\varphi)
={4\epsilon_0{\cal E}r \cos \varphi\over {\pi \sqrt{a^2-r^2}}}
\eqno(5.8)$$
where $\varphi$ measured from the axis of the electric field.
The coefficient $C$ occuring in (2.16) is therefore given by
$$C(s)={2\sqrt{2}\over{\pi}} \epsilon_0 {\cal E}\sqrt{a} \cos \varphi
\eqno(5.9)$$
where $s=a\varphi $ is the distance around the perimeter. 
In two dimensions, the density of states per spin is
$$\nu={m\over{2\pi \hbar^2}}
\ .\eqno(5.10)$$
From (4.8), the total energy transferred to a single electron is
$$\Delta E(t)=\sum_j \delta E_j={2\pi \hbar^2\over{me}}
\sqrt{2\pi \omega\over{v_{\rm F}}}
{\sin \theta\over{\sqrt{\cos \theta}}}
\sum_j C(s_j) \cos \bigl[\omega t_j+\chi_2(\theta)\bigr]
\eqno(5.11)$$
where $t_j=j\tau+t_0$ are the times of the collisions with the
surface, and the sum runs over $N\sim t/\tau$ values of the index $j$. 
The sum in (5.11) is dominated by resonances 
satisfying (1.2). However, in [5] it was shown that when
$\omega \gg \omega_{\rm c}$ (5.11) can be approximated 
by assuming that the 
bounces are independent events. This correctly describes
the average behaviour of the absorption, but not that of the
resonances. The mean squared energy transfer at fixed angle
of incidence $\theta $ is then estimated to be
$$\langle \Delta E^2(t)\rangle \big\vert_\theta =
N {64\pi\hbar^4\epsilon^2{\cal E}^2a\omega\over{m^2e^2v_{\rm F}}}
{\sin^2 \theta \over {\cos \theta}} \langle \cos^2 \phi_j \rangle
\langle \cos^2(\omega t_j+\chi_2(\theta))\rangle$$
$$={8\pi \hbar^4 \epsilon_0^2{\cal E}^2 \omega \over{m^2e^2}}
{\sin^2\theta\over{\cos^2\theta}}t
\equiv A{\sin^2\theta\over{\cos^2\theta}}t
\eqno(5.12)$$
where the final equality defines $A$. 
Now consider how to calculate the phase space
average in (5.4). The average over $\varphi_0$ has 
already been performed. It remains to average over $J$
and $\tau_0$. The quantity being averaged is independent of
$\tau_0$, so that integration over $\tau_0$ gives a contribution
$\tau $. The required average is therefore
$$\langle \Delta E^2 \rangle={
\int dJ\, \tau \langle \Delta E^2\rangle\vert_\theta
\over{\int dJ \, \tau}}$$
$$={\int_0^{\pi/2}d\theta \ 
\cos^2\theta \langle \Delta E^2\rangle\vert_\theta
\over {\int_0^{\pi/2} d\theta\ \cos^2 \theta }}
=At{\int_0^{\pi/2} d\theta \ \sin^2\theta \over
{\int_0^{\pi/2}d\theta \cos^2\theta}}
=At
\ .\eqno(5.13)$$
In the integrable case, the rate of absoprtion of energy is
$dE_{\rm T}/dt=\pi a^2 \nu A$.
The final expression for the rate of absorption of energy
in the integrable case is then
$${dE\over{dt}}={4\pi \hbar^2 \epsilon_0^2 a^2 {\cal E}^2 \omega
\over{me^2}}
\ .\eqno(5.14)$$
It is interesting to note that this expression is independent
of the Fermi energy. It may be written in the form
$${dE_{\rm T}\over{dt}}={(ea{\cal E})^2 (\hbar \omega)
\over {4\pi \hbar E_{\rm R}}}
\eqno(5.15)$$
where $E_{\rm R}=me^4/16\pi^2\epsilon_0^2\hbar^2$ 
is the Rydberg energy, $(ae{\cal E})$ is a measure of the
energy associated with displacement of an electron
across the particle by the electric field, and $\hbar \omega$ is
the photon energy. This result differs by a numerical factor 
from that obtained in [6] using the Thomas-Fermi approximation
for the effective potential.
\par 
If the edge of the disc is rough so that the motion is ergodic, 
the rate of absorption of energy is calculated from  
(5.5) and (5.6). Using (4.8), (5.9) and (5.10), the 
mean squared value of the energy transferred at a single collision is
$$\langle \delta E^2\rangle
={16\pi \hbar^4 \epsilon_0^2{\cal E}^2a\omega
\over{m^2e^2v_{\rm F}}}
\bigg\langle {\sin^2 \theta\over{\cos \theta}}\bigg\rangle
\ .\eqno(5.16)$$
The average over the angle of incidence is
$$\bigg\langle {\sin^2 \theta\over{\cos \theta}}\bigg\rangle
=\int dJ \tau \sin^2\theta/\cos \theta \bigg/
\int dJ \tau
=\int_0^{\pi/2}d\theta \cos \theta \sin^2\theta \bigg/
\int_0^{\pi/2}d\theta \cos^2\theta={4\over {3\pi}}
\ .\eqno(5.17)$$
Also,
$$\langle L \rangle=v_{\rm F}\int dJ\, \tau^2\bigg/\int dJ\, \tau
=2a\int_0^{\pi/2}d\theta \cos^3\theta \bigg/\int_0^{\pi/2}d\theta \,
\cos^2\theta={16a\over{3\pi}}
\ .\eqno(5.18)$$
In the ergodic case, the rate of absorption is
$${dE_{\rm T}\over {dt}}= 
{2\pi \hbar^2 \epsilon_0^2a^2{\cal E}^2\omega\over{me^2}}
\ .\eqno(5.19)$$
This differs by a factor of $1\over 2$ from the integrable case, given
by (5.14).
\gap
\noindent{\sl 5.3 Absorption by conducting spheres}
\gap
The calculation of the energy absorbed proceeds by analogy
with that for conducting discs. The first step is to specify
a convenient set of phase-space coordinates.
In the case of ballistic and specularly reflected electrons
moving in a spherical enclosure, angular momentum is a 
conserved quantity. The following variables can be used
to specify the phase-space coordinates of the electron:
its energy $E$, angular momentum vector ${\bf J}$, and
two angle variables, $\varphi$ and $\varphi'$. In [5], it
was shown that the measure $d\alpha$ for canonical coordinates
is given by
$$d\alpha={\tau\over{J}}\, dE\, d\varphi\, d\varphi'\, d{\bf J}
\ .\eqno(5.20)$$
The energy transferred to an electron bouncing at the surface is
given by equations (4.19) or (4.22). The charge density at the 
surface is that given by classical electrostatics, namely
$$q=3\epsilon_0 {\cal E} \cos \chi\eqno(5.21)$$
where ${\cal E}$ is the amplitude of the external electric field,
and $\chi $ is the polar angle of the point on the surface measured
from the direction of the electric field. 

The rate of absorption of energy is calculated using (5.1) and (5.2).
In the case of integrable motion (specular reflection at the 
surface), the mean squared energy transferred to the 
electron $\langle \Delta E^2\rangle$ is obtained by first 
averaging $\delta E^2$ at fixed angular momentum ${\bf J}$, and then
integrating with respect to angular momentum. The variance of
$\delta E$ at fixed ${\bf J}$ is 
$$\langle \delta E^2 \rangle \vert_{\bf J} =
{9\omega^2\epsilon_0^2{\cal E}^2\over{e^2\nu^2v_{\rm F}^2}}
\cos^2\!\!\chi_0 S^2(\theta) \equiv A \cos^2\chi_0 S^2(\theta)
\eqno(5.22)$$
where $\chi_0$ is the angle between the angular momentum
vector ${\bf J}$ and the direction of the external electric field, and
the final equality defines $A$.
The number of collisions per unit time is $1/\tau$. Treating
the collisions as if they are independent events gives 
$$\langle \Delta^2 E(t)\rangle=
At\int d{\bf J}\ {\tau\over J} {1\over \tau} \cos^2\chi_0 S^2(\theta)
\bigg/
\int d{\bf J} {\tau\over J}
$$
$$=At
\int dJ \int d\chi_0 2\pi J^2 \sin\chi_0 {1\over J} 
S^2(\theta)\cos^2\chi_0
\bigg/
\int dJ 4\pi J^2 {\tau\over J}
$$
$$={3\epsilon_0^2\omega^2{\cal E}^2\over{2e^2\nu^2 a v_{\rm F}}}
\int_0^{\pi/2} d\theta\ \cos\theta \sin\theta S^2(\theta)
\bigg/
\int_0^{\pi/2} d\theta\ \cos^2\theta \sin\theta
\eqno(5.23)$$
The rate of absorption of energy by the spherical particle is 
therefore
$${dE\over{dt}}={6\pi \epsilon_0^2 a^2 {\cal E}^2\omega^2 
\over{e^2\nu v_{\rm F}}}{\cal F}(\Gamma)
\eqno(5.24)$$
where
$${\cal F}(\Gamma)=\int_1^\infty dx {1\over {x}}  
{\big\vert 1-\Gamma x^2g(x)\big\vert^2
\over{
(1+\Gamma x^2)^2\big\vert 1+g(x)\big\vert^2
}}
\ .\eqno(5.25)$$
The integral ${\cal F}(\Gamma)$ diverges logarithmically 
as $\Gamma \to 0$: for small $\Gamma$, 
${\cal F}\sim K-{1\over 2}\log_{\rm e}(\Gamma)$, where $K$ 
is a constant. In the limit $\Gamma \to \infty$, ${\cal F}$ 
approaches a finite limit. Some values of ${\cal F}(\Gamma )$ obtained
by numerical integration are given in table 1. 
A clearer understanding of (5.24) is obtained by expressing
it in terms of ratios of energies: two equivalent forms are 
$${dE_{\rm T}\over {dt}}=
{3\pi\over 8}{(ae{\cal E})^2 (\hbar \omega)^2
\over{\hbar E_{\rm F}E_{\rm R}}}
{\cal F}(\Gamma)
=3\pi^3{(ae{\cal E})^2 (\hbar \omega)^2
\over{\hbar E_{\rm R}^2}}
\Gamma^2{\cal F}(\Gamma)
\eqno(5.26)$$
\par
In the case where the surface of the spherical particle is rough,
the electron motion is ergodic, and the rate of absorption 
is calculated via the microcanonical average
of $\delta E^2$, using (5.5) and (5.6). A complicating feature 
is that the charge density concentrates on prominences
of a rough surface. In [5] a simplified model was discussed, in which
a fraction $\eta $ of the surface comprises high plateaus, with the
the charge density is increased by a factor $1/\eta$, and the remainder
of the surface is un-charged. According to this model, $\delta E^2$
is increased by a factor of $1/\eta^2$ for a fraction $\eta $ of 
collisions. The required average is
$$\langle \delta E^2 \rangle =
{2A\over \eta}\int d{\bf J}\ {\tau\over J}\cos^2\chi S^2(\theta)
\bigg/
\int d{\bf J}\ {\tau\over J}
$$
$$={2A\over \eta}
\int dJ \int d\chi 2\pi J^2 \sin\chi {\tau \over J}\cos^2\chi  S^2(\theta)
\bigg/
\int dJ 4\pi J^2 {\tau\over J}
$$
$$={2A\over \eta}\int_0^{\pi/2}d\theta \ \cos^2\theta \sin\theta S^2(\theta)
\eqno(5.27)$$
where $A$ is the factor defined in (5.22).
The average distance between bounces is
$$\langle L\rangle=v_{\rm F}\langle \tau \rangle={3a\over 2}
\ .\eqno(5.28)$$
The rate of absorption is then found to be
$${dE_{\rm T}\over{dt}}={16\pi \epsilon_0^2a^2 {\cal E}^2\omega^2
\over{\eta e^2\nu v_{\rm F}}}{\cal G}(\Gamma)\eqno(5.29)$$
where
$${\cal G}(\Gamma)=\int_1^\infty dx {1\over{x^2}}
{\vert 1-\Gamma x^2g(x)\vert^2
\over{(1+\Gamma x^2)^2\big\vert(1+g(x)\big\vert^2}}
\ .\eqno(5.30)$$
The function ${\cal G}(\Gamma )$ appraoches finite limits
as $\Gamma \to \pm \infty$. Values obtained by 
numerical integration are given in table 1.
\gap\gap\gap
%
%
%
%%%%%%%%%%%%%%%%%%%%%%%%%%%%%%%%%%%%%%%%%%%%%%%%%%%%%%%%%%%%%%%%%%%%%
%
\noindent{\bf 6. Concluding remarks}
\gap
The principal new results in this paper were the solutions
of the simplified RPA equations for the effective potential
$\phi(z)$ (section 3), and the calculation of the energy transferred to
an electron rebounding from the surface (section 4). They improve upon
the earlier analysis in [5] and [6], which used a Thomas-Fermi
approximation for the effective potential, rather than the 
random phase approximation (RPA). In section 5 these results were used 
to obtain an improved estimate for the absorption of radiation
by small particles, at frequencies small compared to the plasma
frequency.

It is desirable to consider the limitations of some of the 
approximations which have been employed. The RPA prescription
itself is an uncontrolled approximation, but it is expected
to work well at high electron densities (small $r_{\rm s}/a_0$,
or equivalently at small values of the parameter $\Gamma $
defined in (4.21)). 

The image source method simplifies the calculation of the 
polarisability in two ways. Firstly, it treats the surface
as if it were flat, and ignores contributions arising from
reflections at more distant parts of the surface. These 
contributions are assumed to have relatively smaller 
amplitude and to combine incoherently, but their effect is
very hard to quantify (except for slabs or strips
with flat parallel faces). A second, and more significant,
defect of the image source approximation is that it 
ignores interference effects between the direct and 
reflected paths. These are most significant when the
amplitudes for both paths are comparable: this happens
when either ${\bf r}$ or ${\bf r}'$ is close to the
surface, and results in the polarisability vanishing as
either point approaches the surface. The polrisability
will differ from (2.4) by interference terms which oscillate
with a wavenumber comaparble with the Fermi wavenumber, $k_{\rm F}$.

The semiclasical RPA approximation scheme makes the further
assumption that the charge density has its classical distribution.
In three dimensions the semiclassical RPA potential was compared with
the image source approximation. The former has a delta function
at the boundary, whereas the latter has a finite slope at $z=0$,
corresponding to a rapid initial decay with length scale
$\lambda_{\rm s}=\sqrt {\epsilon_0/e^2\nu}$ equal to the Thomas-Fermi
screening length. 

The parameter $\Gamma $ which appears in the image source theory
for the energy transfer on reflection is small when $r_{\rm s}/a_0$
is small. In this limit the ratio of the
Thomas-Fermi screening length to the Fermi wavelength,
$\lambda_{\rm s}/\lambda_{\rm F}$ is large. When 
$\lambda_{\rm s}$ is small compared to the Fermi wavelength,
the approximations used here are expected to fail, because the 
polarisability approaches zero within $\lambda_{\rm F}$ of the surface,
whereas the charge is expected to accumulate with a layer of
thickness $\lambda_{\rm s}$. These considerations indicate that 
the theory is more accurate when $\Gamma $ is small, which 
corresponds to the case of good metals with a high
density of conduction electrons. It is even possible
that the results might be asymptotic to the results of
an exact implementation of the RPA equations in the limit
$\Gamma \to 0$.

In summary, the results presented in this paper represent an
approximation scheme which leads to analytic expressions
for the effective potential $\phi(z)$, the energy transferred
to an electron rebounding from the surface $\delta E(\theta)$
and the rate of energy absorption by a small particle due
to an oscilating electric field. The formulae
are expected to be a good approximation to the full RPA equations
when $\Gamma $ is small, but it would be desirable to benchmark
them against a numerical evaluation of the full RPA scheme.
\gap\gap\gap
%
%
%
%%%%%%%%%%%%%%%%%%%%%%%%%%%%%%%%%%%%%%%%%%%%%%%%%%%%%%%%%%%%%%%%%%%%%
%
\vfill
\eject
{\indentoff \bf References}
\gap
\indentoff
\gap
[1] A. L. Fetter and J. D. Walecka,
{\sl Quantum Theory of Many-particle Systems},
New York: McGraw Hill, (1971).
\gap\gap
[2] M. Wilkinson and B. Mehlig, {\sl J. Phys.: Condens. 
Matter}, in press.
\gap\gap
[3] L. D. Landau and E.  M. Lifshitz, {\sl Electrodynamics
of Continuous Media (Landau and Lifshitz Course
of Theoretical Physics 8)}, Oxford: Pergamon (1958).
\gap\gap
\ref {4}{B. Mehlig and M. Wilkinson}{J. Phys.: Condens. Matter}
{9}{3277-90}{1997}
\gap
\ref {5}{E. J. Austin and M. Wilkinson}{J. Phys.: Condens. Matter}
{5}{8461-8484}{1993}
\gap
\ref {6}{M. Wilkinson and E. J. Austin}{J. Phys.: Condens. Matter}
{6}{4153-4166}{1994}
\gap
\ref {7}{M. Wilkinson and B. Mehlig}{Eur. J. Phys.}{1B}{397-8}{1998}
\vskip 4 true cm
\noindent{\bf Figure captions}
\gap
\noindent Figure 1. Schematic plot, illustrating the three
components of the effective potential.
\gap
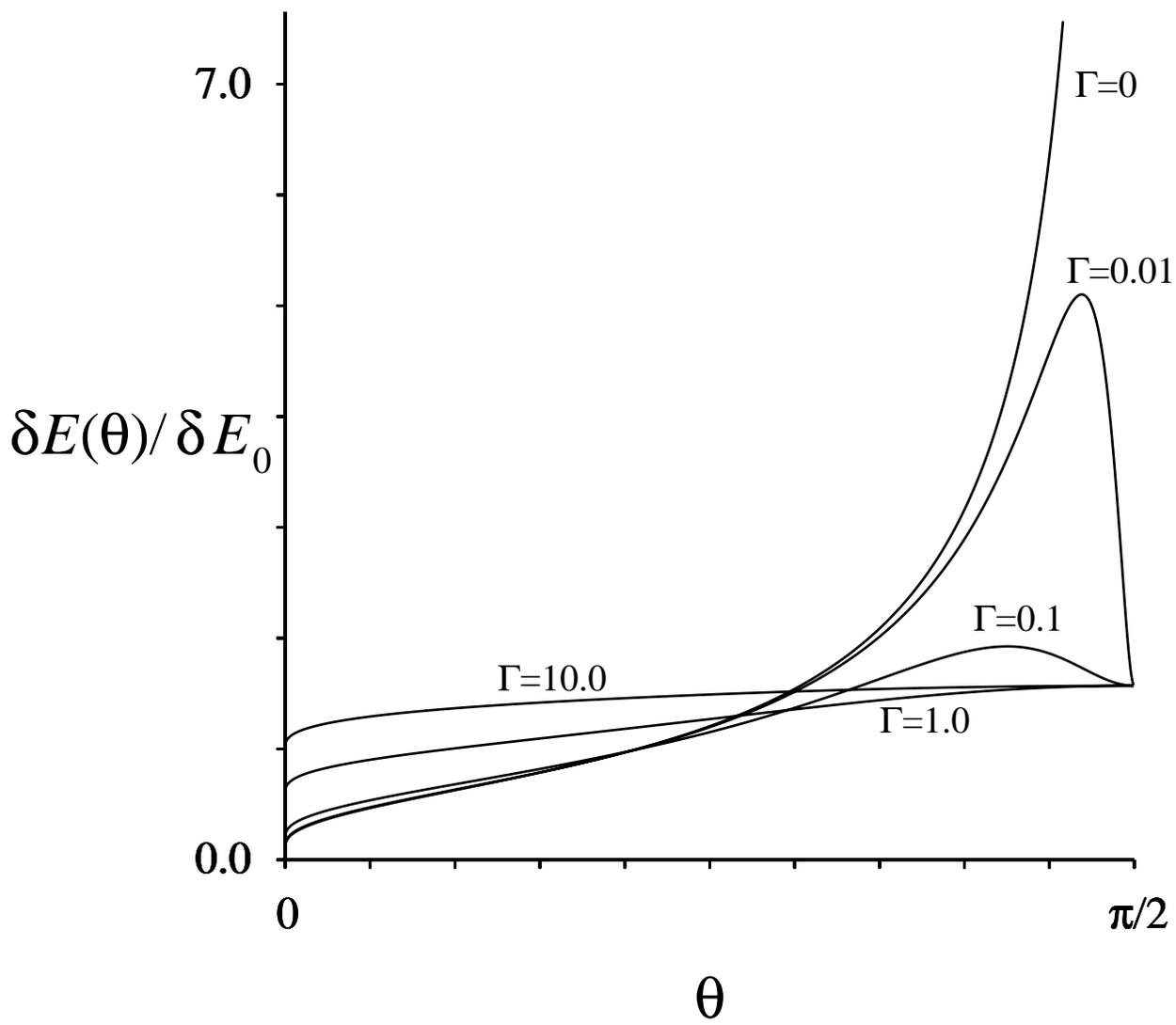
\noindent Figure 2. The energy $\delta E$ transferred to an electron
rebounding from the surface, as a function of angle of incidence
$\theta$. The scale energy is $\delta E_0=2q\omega/e\nu v_{\rm F}$. 
This function depends upon the material-dependent parameter
$\Gamma \propto r_{\rm s}/a_0$.
\vskip 4 true cm
\noindent {\bf Table caption}
\gap
\noindent Table 1. Values of the functions ${\cal F}(\Gamma)$ 
and ${\cal G}(\Gamma)$, defined by (5.25) and (5.30), obtained 
by numerical integration.
\gap\gap\gap
\vfill
\eject
\end